\documentclass[aps,prb,superscriptaddress,twocolumn,showpacs]{revtex4-2}

\usepackage[utf8]{inputenc}
\usepackage[english]{babel}
\usepackage{graphicx} 

\makeatletter\AtBeginDocument{\let\@elt\relax}\makeatother

\usepackage{amsmath}
\usepackage{amssymb}
\usepackage{xcolor}
\usepackage{color} 
\usepackage[abs]{overpic}
\usepackage{epstopdf}
\usepackage{soul}

\usepackage[normalem]{ulem}

\begin{document}

\title{Dynamics of inhomogeneous spin ensembles with all-to-all interactions: \\ breaking  permutational invariance}

\author{ Fernando Iemini}
\affiliation{Instituto de F\'isica, Universidade Federal Fluminense, 24210-346 Niter\'oi, Brazil}
\affiliation{Institute for Physics, Johannes Gutenberg University Mainz, D-55099 Mainz, Germany}

\author{ Darrick Chang}
\affiliation{ICFO-Institut de Ciencies Fotoniques, The Barcelona Institute of
Science and Technology, 08860 Castelldefels (Barcelona), Spain}
\affiliation{ICREA-Institucio Catalana de Recerca i Estudis Avancats, 08010 Barcelona, Spain}

\author{ Jamir Marino}
\affiliation{Institute for Physics, Johannes Gutenberg University Mainz, D-55099 Mainz, Germany}

\date{\today}

\begin{abstract}

We investigate the consequences of introducing non-uniform initial conditions in the dynamics of spin ensembles characterized by all-to-all interactions. Specifically, our study involves the preparation of a set of semi-classical spin ensembles with varying orientations. Through this setup, we explore the influence of such non-uniform initial states on the disruption of permutational invariance.
Comparing this approach to the traditional scenario of initializing with spins uniformly aligned, we find that the dynamics of the spin ensemble now spans a more expansive effective Hilbert space. This enlargement arises due to the inclusion of off-diagonal coherences between distinct total angular momentum subspaces—an aspect typically absent in conventional treatments of all-to-all spin dynamics.
Conceptually, the dynamic evolution can be understood as a composite of multiple homogeneous sub-ensembles navigating through constrained subspaces. Notably, observables that are sensitive to the non-uniformity of initial conditions exhibit discernible signatures of these off-diagonal coherences.
We adopt this fresh perspective to reexamine the relaxation phenomena exhibited by the Dicke model, as well as a prototypical example of a boundary time crystal. Intriguingly, ensembles initialized with inhomogeneous initial conditions can show  distinctive behaviors when contrasted with canonical instances of collective dynamics. These behaviors encompass the emergence of novel gapless excitations, the manifestation of limit-cycles featuring dressed frequencies due to superradiance, instances of frequency locking or beating synchronizations, and even the introduction of ``extra'' dimensions within the dynamics.
In closing, we provide a brief overview of the potential implications of our findings in the context of modern cavity quantum electrodynamics (QED) platforms. 


\end{abstract}

\maketitle

\section{Introduction}

Non-equilibrium dynamics in systems with long-range interactions have been a subject of growing interest in recent years, strongly stimulated by the experimental progresses for their simulations \cite{Dimer_2007,Puri_2017,Zhang_2017,Angerer_2018,Henriet_2020}. Their eminent collective character can support novel phenomena (in contrast to their short-range interacting counterparts), including new bounds on information spreading \cite{Chen_2023} with super-luminal correlations \cite{Richerme_2014,Jurcevic_2014}, different critical behaviors \cite{Defenu_review,Jamir_2022} and routes to thermalization \cite{Ueda_review_therm}, among others \cite{Silvia_review,Jreview}. Understanding the non-equilibrium dynamics of such systems is of paramount importance, not only for unraveling fundamental aspects of many-body systems but also for exploring their potential applications 
in quantum technologies \cite{Genes_review2022}.

 All-to-all interactions arise as the limiting case where the range of interactions spans over distances larger than the size of the system. Albeit a physically idealized regime, it may still capture 
 main properties of systems whose range of interactions are sufficiently long as compared to their effective dimensionality \cite{Defenu_review}.
A striking phenomena under this regime is the well-known superradiance, first studied by Dicke \cite{Dicke_1954} in order to describe the emission of light by a large ensemble of $N$ atoms prepared in their excited state. Considering that the decay of the atoms occur in a collective fashion - therefore allowing coherence and constructive interferences - the emission process can trigger a burst of radiation with a rate proportional to $N^2$ (as opposed to $N$ for independent atoms).
Different forms of superradiance also occur for an ensemble of atoms coupled to a single cavity mode which mediates the all-to-all interactions among the atoms \cite{Hepp_1973,Dimer_2007,Kirton_2019}. In this case the steady state of the coupled spin-cavity system can stabilize exhibiting a macroscopic occupation for its photonic field.
In fact, a multitide of interesting phenomena are supported in fully connected spin ensembles \cite{Genes_review2022}.
 Particular attention has been devoted recently to their capability to support time crystals both in closed \cite{Russomanno_2017,lorenzo2017quantum, Surace2019,Nurwantoro2019,Lyu2020,Pizzi2021,Collado2021,Manuel2022,Morita2006,Xiaoqin2021} and open systems \cite{Iemini2018_btc,Gong2018,Wang2018,Tucker2018,Lesanovsky2019,Zhu2019,Buca2019,Jiasen2023,Prazeres2021,Piccito2021,Antonio2022,Kongkhambut2022,Kessler2021,Tuquero2022,Vinjanampathy2022,Passarelli2022,Buca2022,Krishna2023,Mattes2023,Peng2023,Nakanishi2023,Chen2023}, where long-range interactions prevents thermalization and moreover stabilize persistent collective dynamics which breaks the time-translation symmetry of the system.

The usual approach in the study of such systems assumes the spins in the ensemble as permutationally invariant (PI) \cite{Chase2008,Shammah2018,Baragiola2010,Shammah2017,Sarkar1987,Xu2013,Gegg2018,Hartmann2016,Martini2001,Huybrechts2020,Bastin2023}. Specifically, given that the spins are identically prepared, the dynamics preserves such a symmetry allowing a considerably simplification for its analysis. The exponentially large Hilbert space can be reduced to an effective polynomial size in order to fully represent the density matrix \cite{Chase2008,Shammah2018,Sarkar1987,Hartmann2016,Xu2013}.

In this work we explicitly break the permutation symmetry in an all-to-all interacting ensemble of spins from its initial preparation. The inhomogeneous ensemble therefore accesses larger portions of the Hilbert space, but importantly still with a constrained dynamics imposed by the nature of their interactions. Specifically, we show that the system is equivalent to a set of $M < N$ permutationally invariant ensembles which evolve as a fully connected system and within a correlated dynamics conserving their angular momentum angles. We explore this scenarion by revisiting   the Dicke model and a boundary time crystal.
In the Dicke model we observe the emergence of new gapless excitations in the off-diagonal subspaces, inducing the appearance of limit-cycles with dressed frequencies by the expectation value of the magnetization in the symmetry broken phase, and simple Rabi oscillatory dynamics for the normal phase. We show that they are both direct consequences of  permutational symmetry breaking. Remarkably,   boundary time crystal phases can add an extra dimension to the long-time dynamics of the inhomogeneously prepared spin ensembles. We observe a plethora of effects including frequency locking  synchronization between observables which can or cannot detect the initial inhomogeneity, as well as beating dynamics among them. 
Interestingly, the study of the dynamics away from the PI in all-to-all systems has also been recently under scrutinity in the context of scrambling and entanglement growth \cite{Zihao2023}, denoted as dynamics in the \textit{deep Hilbert space}, with the emergence of new super-exponential growth, or in (unitarily equivalent) unconventional Dicke models \cite{Farokh_2023} with the emergence of new multistabilities and nonequilibrium dynamics.

The manuscript is organized as follows. In Sec.\eqref{sec.HS.decomposition} we introduce the class of Lindbladian systems we study, and its Hilbert space structure with conserved quantities and symmetries. In Sec.\eqref{sec.M2.subens} we discuss the effects of inhomogeneities in the initial state, with focus on the distribution of the diagonal and off-diagonal density matrix onto the conserved subspaces of the dynamics and the role of observables to capture such effects. We show the equivalence of the inhomogeneous ensemble to $M$ PI subensembles with a constrained dynamics. In Sec.\eqref{sec.Dicke.model} and Sec.\eqref{sec.btc} we analyse different models within this perspective, namely the Dicke model and a boundary time crystal. We show the emergence of  peculiar dynamic behaviors specific to the presence of inhomogeneity, i.e., to the breaking of permutation invariance in the system. Sec.\eqref{sec.conclusion} is devoted to our conclusions and perspectives.

\section{Hilbert space decomposition}
\label{sec.HS.decomposition}

We consider collective all-to-all dissipative dynamics for a spin ensemble, driven by Lindbaldian dynamics
 $d \hat \rho/dt =  \mathcal{L}[\hat \rho] $, as follows,
\begin{equation}\label{eq.spin.gen.Lindbladian}
 \mathcal{L}[\hat \rho] = -i[\hat H,\hat \rho] + \sum_{\vec{\alpha},\vec{\beta}}  \Gamma_{\vec{\alpha},\vec{\beta}} \left( \hat S^{\vec{\alpha}} \hat \rho (\hat S^{\vec{\beta}})^\dagger - \frac{1}{2}\{(\hat S^{\vec{\beta}})^\dagger \hat S^{\vec{\alpha}}, \hat \rho\} \right)
\end{equation}
with coherent Hamiltonian described by,
 \begin{equation}
\hat H = \sum_{p=1}^\infty \frac{1}{N^{p-1}}\sum_{\{ \vec{\alpha}, ||\vec{\alpha}||=p \}} \omega_{\vec{\alpha}} \hat S^{\vec{\alpha}},
\end{equation}
where $\vec{\alpha} = (\alpha_1, ...,\alpha_p)$ is a vector with $p$ elements indicating the direction $\alpha_k = x,y,z$ of the collective spins $\hat S^{\vec{\alpha}} = \hat S^{\alpha_1}...\, \hat S^{\alpha_p}$, 
with $\hat S^\alpha = (1/N)\sum_{i=1}^N \hat \sigma_i^\alpha/2$  the collective spin operators, $N$ is the total number of spins in the ensemble and  
$\sigma_i^\alpha$ the Pauli spin operator for the $i$th spin in the ensemble. The collective operators inherit the algebra of their microscopic spins, satisfying the $SU(2)$ algebra with commutation relations given by $[\hat S^\alpha, \hat S^\beta] = i \epsilon^{\alpha \beta \gamma} \hat S^\gamma$. 

\begin{figure}
 \includegraphics[width = 0.9 \linewidth]{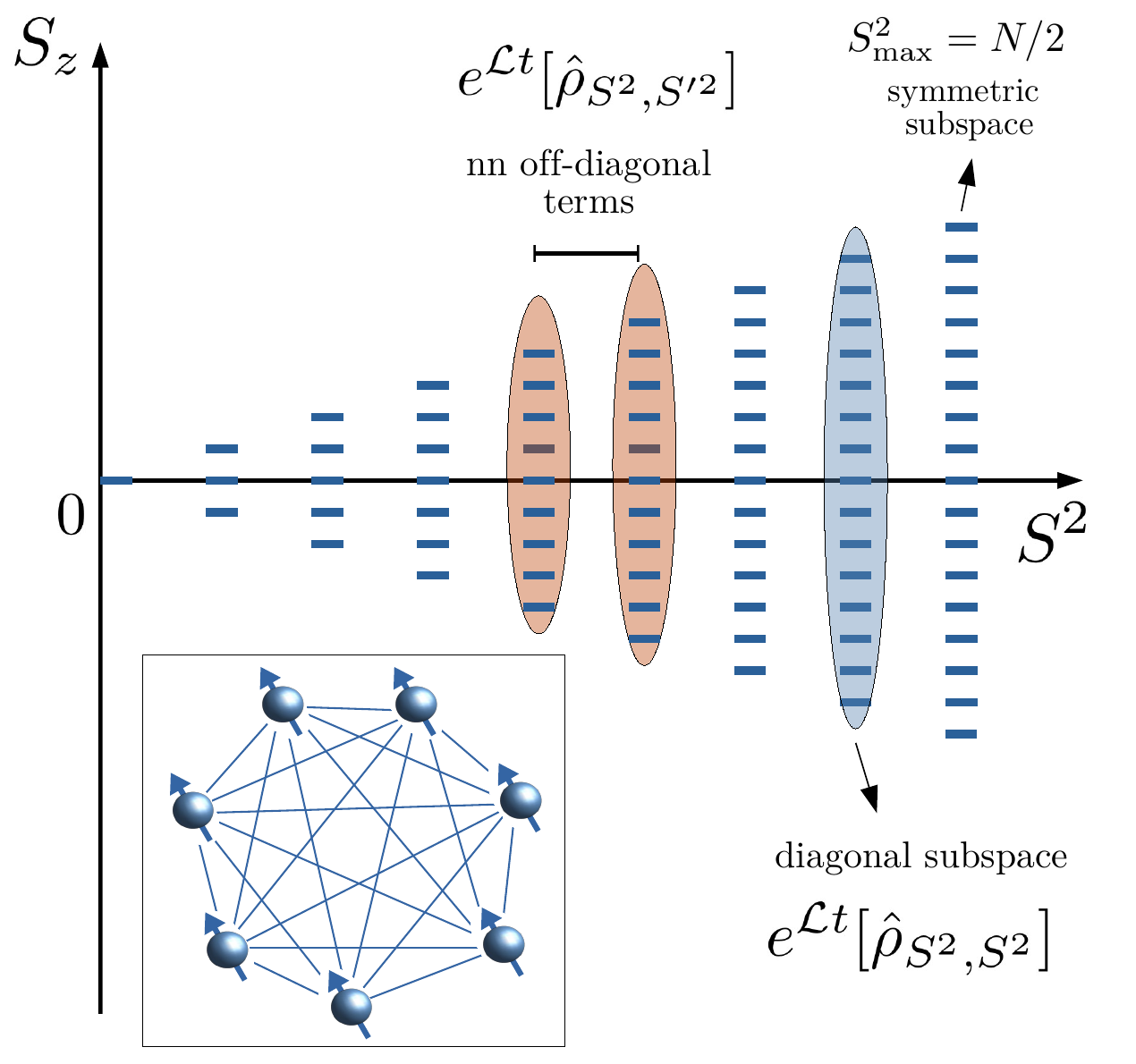}
 \caption{ Hilbert space decomposition for permutationally invariant Dicke states. The states are fully described by quantum numbers of total angular momentum $S^2$ and collective magnetization $S_z$. The maximum angular momentum subspace $S^2 = N/2$ corresponds to the symmetric subspace. In systems with all-to-all interactions (inset panel) the total angular momentum is preserved, and the dynamics decouples among the diagonal ($e^{\mathcal{L}t}[ \hat  \rho_{S^2,S^2}]$) and off-diagonal ($e^{\mathcal{L}t}[\hat \rho_{S^2,S'^2}]$, $S^2 \neq S'^2$) terms of these quantum number subspaces.
 }
 \label{fig.HS.decomposition}
\end{figure}

These dynamics can manifest in  different contexts. We first recall that it corresponds to an effective Markovian dynamics for the spins, obtained by tracing out the environment degrees of freedom under the assumption of a memory-less and product state environment -  i.e, Born-Markov approximation \cite{Petruccione_book}. At a purely theoretical level it always possible to define a system-environment Hamiltonian whose effective dynamics is described by the given Lindbladian \cite{Chebotarev_1997,Gregoratti_2001,Gough_2015}. From an experimental perspective, the implementation of effective all-to-all interactions have been demonstrated on many different state-of-art platforms, such as cavity QED systems \cite{Dimer_2007}, artificial qubits in superconducting circuits \cite{Puri_2017}, trapped ions \cite{Zhang_2017}, color defects in diamonds \cite{Angerer_2018}, and Rydberg atoms \cite{Henriet_2020}. Later in the manuscript we  focus on a possible implementation of our ideas in  cavity QED systems where several atomic ensembles can interact via one or multiple common cavity modes.

An important property of these systems arise from the  collective all-to-all nature of the operators, which conserve the total spin angular momentum, $\hat S^2 = (\hat S^x)^2 + (\hat S^y)^2 + (\hat S^z)^2$. Therefore we can represent the state of the system using the quantum numbers for the total angular momentum and collective magnetization along a direction, namely, the Dicke states $\{|S^2,S_z,i\rangle\}$ \cite{Dicke_1954}. The quantum numbers range from
$S^2= 0,1,..., N/2$ for the total angular momentum, $S_z = -S^2,...,+S^2$ for the spin magnetization  and $i = 1,...,d_N^{S^2}$ for their irreducible representations. The algebra of collective operators on such space is given by, 
\begin{eqnarray}
 \hat S^2 |S^2,S_z,i\rangle &=& S^2(S^2+1) |S^2,S_z,i\rangle \nonumber \\
 \hat S^z |S^2,S_z,i\rangle &=& S_z |S^2,S_z,i\rangle  \label{eq.algebra.coll.spins} \\
 \hat S_{\pm} |S^2,S_z,i \rangle &=& \sqrt{(S^2\mp S_z)(S^2\pm S_z+1)} |S^2,S_z\pm 1,i\rangle \nonumber
\end{eqnarray}
with $\hat S_{\pm} = \hat S^{x}\pm i \hat S^{y}$. On its simplest form, the Dicke states for the maximun angular momentum $S^2=N/2$ has no irredutible degeneracy ($d_N^{S^2}=1$) and corresponds to the symmetric subspace among the microscopic spins,
\begin{equation}
 |S^2=N/2,S_z \rangle = \frac{1}{\sqrt{\mathcal{N}}} \sum_{\pi \in \mathcal{S}_N} P_{\pi} \left( |0\rangle^{N/2 - S_z} 
 |1 \rangle^{S_z} \right) 
\end{equation}
with $P_{\pi}$ the permutation operator among the spins,  $\mathcal{S}_N$ the symmetric group and $\mathcal{N}$ the normalization constant. While the symmetric subspace is easly represented in terms of its spin tensor decomposition, Dicke states with lower total angular momentum have rather intrincate structures among its microscopic spins given by their corresponding Clebsch-Gordan coefficients. A general density matrix is therefore described in this basis as,
\begin{equation}
 \hat \rho = \sum_{(S^2, S_z,i),(S'^2,S'_z,i')} c_{S^2, S_z,i}^{S'^2, S'_z,i'} |S^2,S_z,i\rangle \langle S'^2, S'_z,i'|,
\end{equation}

\begin{figure}
 \includegraphics[width = 1 \linewidth]{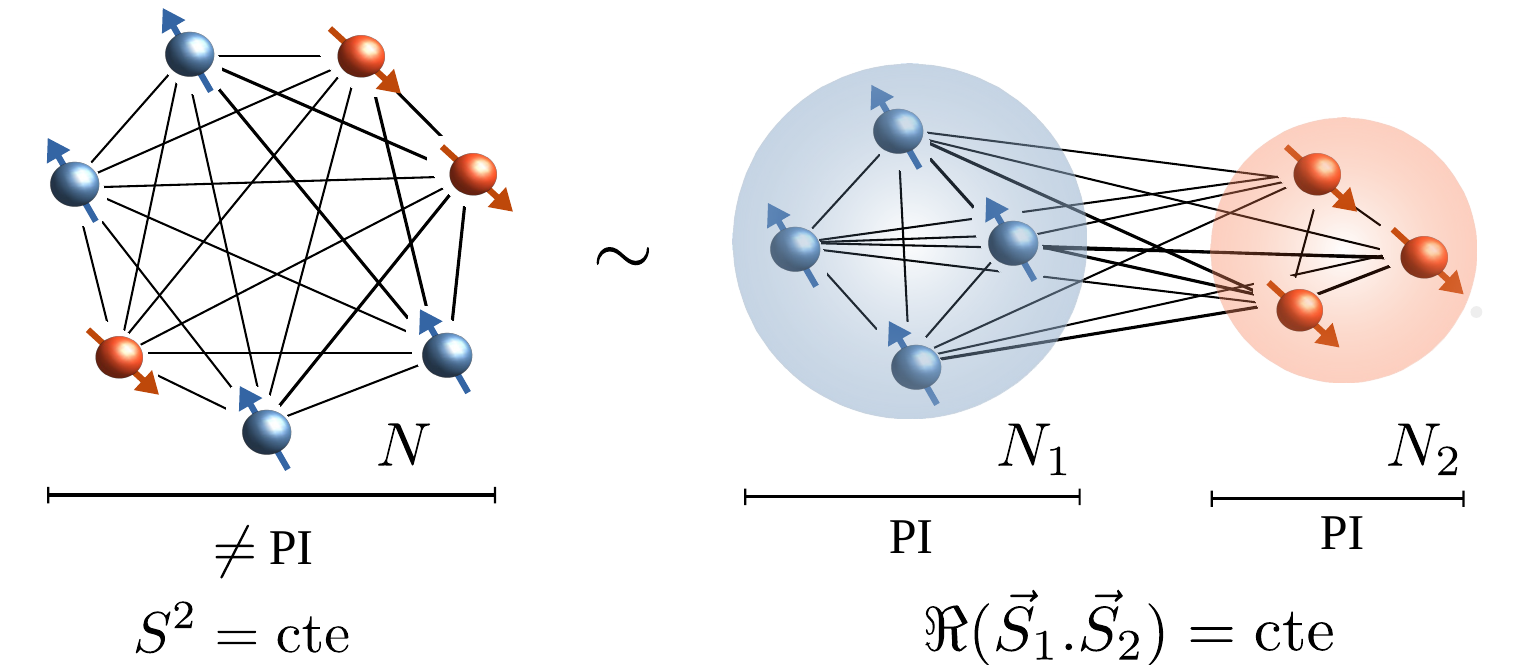}
 \caption{ A general spin ensemble with no permutation symmetry  can always be decomposed into $M$ smaller subensembles with a constrained dynamics, which separately recover the symmetry. The figure illustrates the $M=2$ case, for which a system with $N$ spins with no PI is equivalent to two PI subensembles with $N_1+N_2=N$ 
 with an underlying dynamics conserving the relative angle between their total angular momentum, $\Re( \vec{S}_1 . \vec{S}_2)$.
 }
 \label{fig.M2.ensembles}
\end{figure}

Usual approaches consider permutationally invariant states,  e.g., assuming the $N$ spins are initially prepared in identical states, and recalling that   Lindbladian dynamics preserves such a symmetry. Permutationally invariant states are described by,
\begin{equation}
 \hat \rho = \sum_{\hat \pi \in \mathcal{S}} \hat \pi \hat \rho \hat \pi
\end{equation}
with $\mathcal{S}$ the symmetric permutation group and the sum covering all possible permutations among the spins.
The simplification on such systems relies most on the fact that the irredutible representations of the Dicke states become redudant. In this case the system is fully described by $(S^2,S_z)$ quantum numbers where we can integrate all i'th indexes, describing an effective collective state $|S^2,S_z\rangle$ \cite{Chase_2008} - see Fig.\eqref{fig.HS.decomposition}. Equivalently, one may define effective coefficients 
$c_{S^2, S_z}^{S'^2, S'_z} = \sum_{i,i'} c_{S^2, S_z,i}^{S'^2, S'_z,i'}$ contracting the irredutible indexes. Moreover, permutationally invariant states have another crucial property that allows a simpler description, namely, the absence of coherence among the angular momentum subspaces, 
$c_{S^2, S_z}^{S'^2, S'_z} = \delta_{S^2,S'^2} \, c_{S^2, S_z}^{S'^2, S'_z}$. The state is block diagonal among such subspaces, and its effective Hilbert space dimension scales cubically with the system size, $d_{\rm{PI},N} \sim ~ \mathcal{O}(N^3)$.

In this work we break the permutational invariance of the initial state, considering inhomogeneity among the spins. In this case the density matrix displays coherence among  different angular momentum subspaces with non-negligible effects, as we discuss. Due to the all-to-all nature of the couplings and its $S^2$ conservation, the full Lindbladian dynamics of the system can be partitioned among all different pairs of angular momentum quantum numbers, $\hat \rho_{S^2,S'^2} = \hat P_{S^2} \hat \rho \hat P_{S'^2}$ with  $\hat P_{S^2}$ the projector operator on the subspace with fixed $S^2$ angular momentum. The full density matrix dynamics is given by,
\begin{equation}
 \hat \rho(t) = \sum_{S^2,S'^2} \hat \rho_{S^2,S'^2}(t),
\end{equation}
with $ \hat \rho_{S^2,S'^2}(t) = e^{\mathcal{L}t}[\hat \rho_{S^2,S'^2}(0)]$. From the algebra of collective operators (Eq.\eqref{eq.algebra.coll.spins}) we see that the diagonal subspaces ($S^2=S'^2$) evolve similar to each other apart from a re-scaling of the effective number of spins in the system, $N_{\rm{eff},S^2} \sim S^2$. In other words, the effective number of spins participating on the dynamics depends on the angular momentum of the subspace. While the diagonal terms have such simpler interpretation,
off-diagonal terms do not have a simple description since these depend on left and right acting collective operators, each one having its own and different effective number of spins. Neverthless, since such off-diagonal evolves by decoherence dynamics, one expects in general that such terms vanish in the long time limit, with the non-equilibrium steady state (NESS) of the system described by a mixture of diagonal terms only.

\section{$M=2$ subensembles}
\label{sec.M2.subens}

\begin{figure}
\includegraphics[width = 0.9 \linewidth]{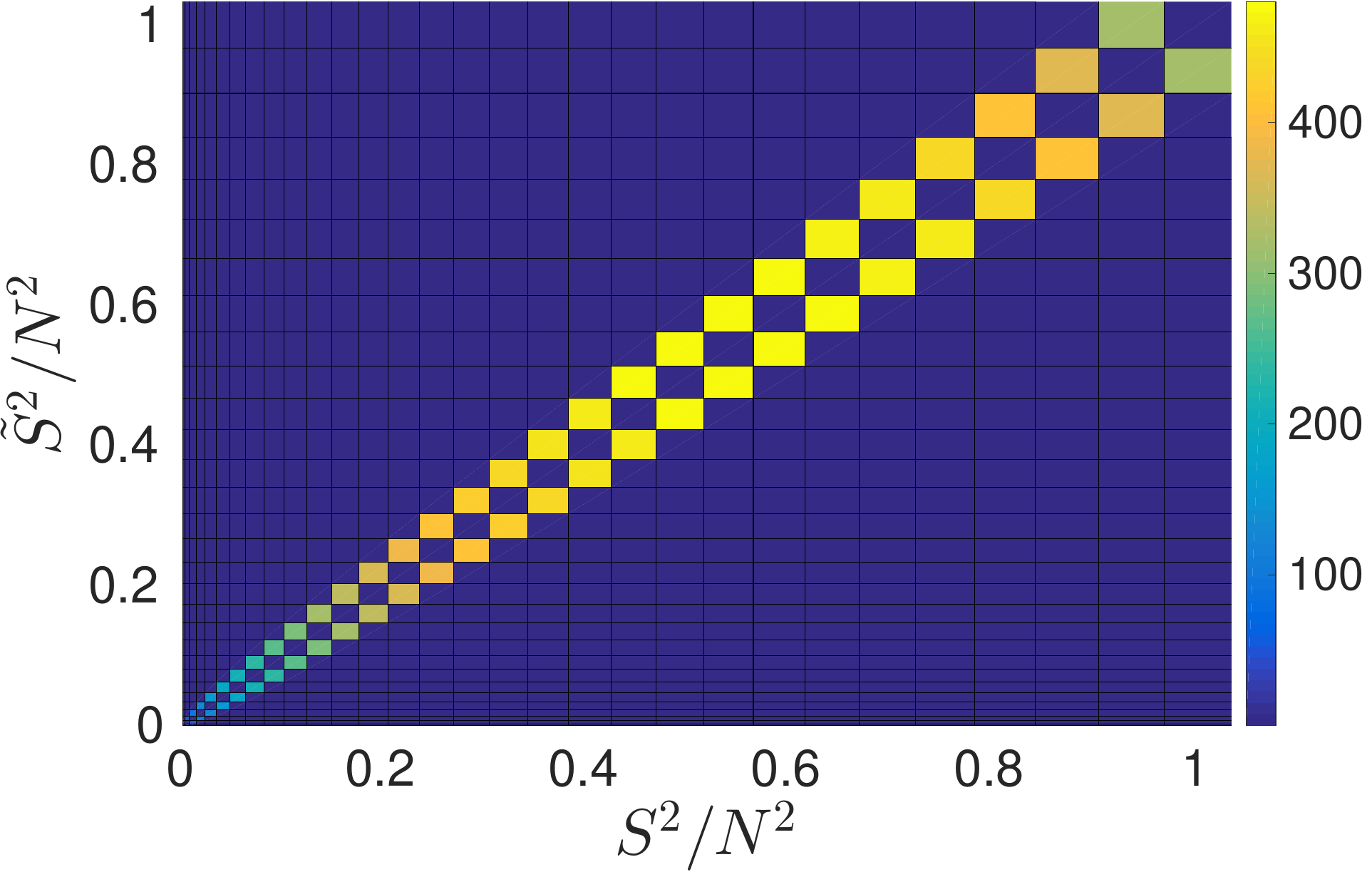}
 \caption{  Distribution of the antisymmetric  magnetization $\hat  O_{\mathcal{A},\alpha=x,y}$ along the different $S^2$ subspaces, highlighting their overlap between nearest-neighbor subspaces. Specifically, we compute the overlap between the $S^2$ and $\tilde{S}^2$ subspaces defined by $ \sum_{S_z,\tilde{S}_z} |\langle S^2,S_z| \hat  O_{\mathcal{A},\alpha=x,y} |\tilde{S}^2,\tilde{S}_z\rangle $.
 The antisymmetric observables here  are defined for $M=2$ spin ensembles, each with $N_{1(2)}=30$ spins. The results for $\hat  O_{\mathcal{A},z}$ are qualitatively similar (not shown).
 }
 \label{fig.antiss.obs.distrib}
\end{figure}

\begin{figure*}
\includegraphics[width = 0.32 \linewidth]{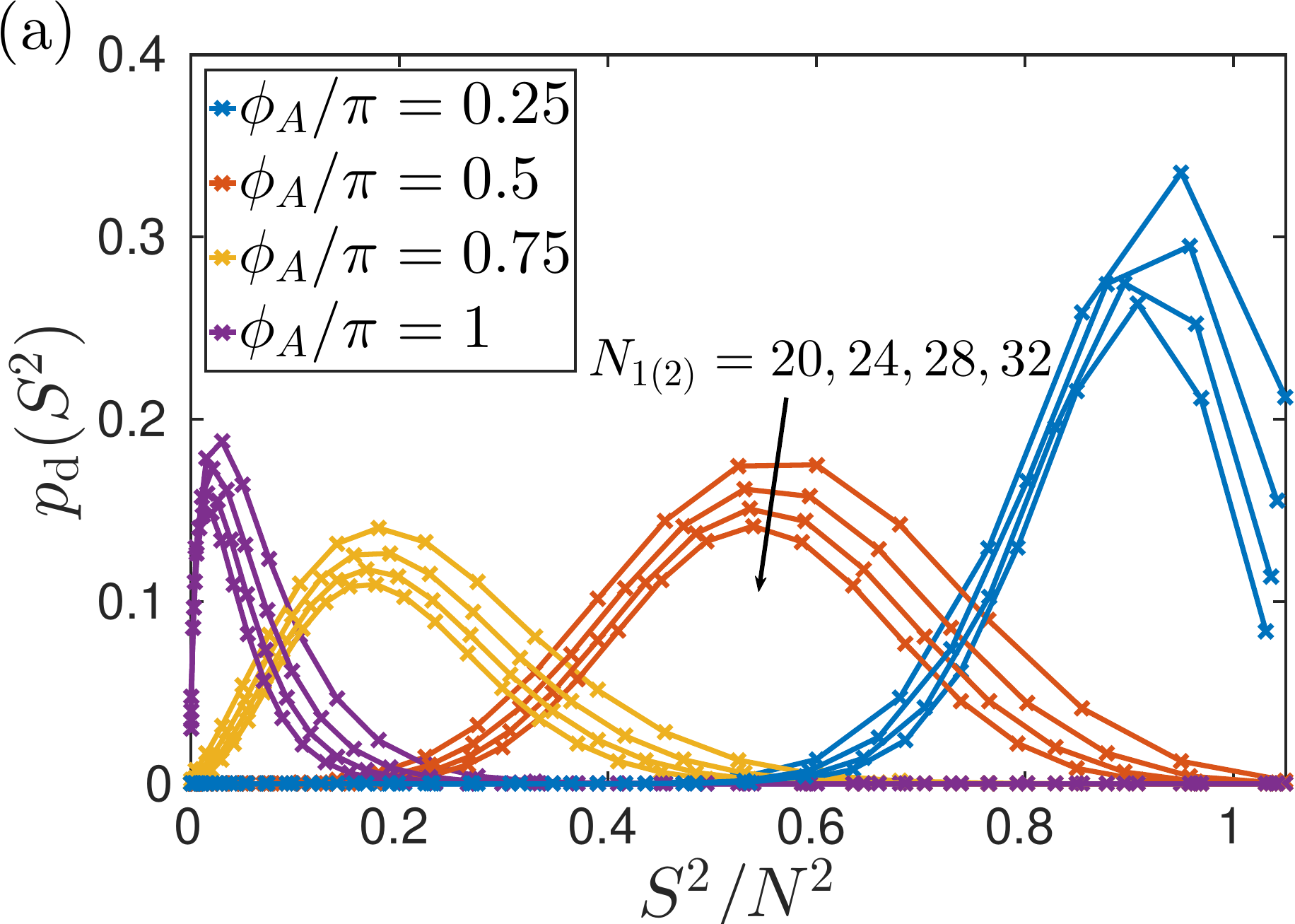}
\includegraphics[width = 0.34 \linewidth]{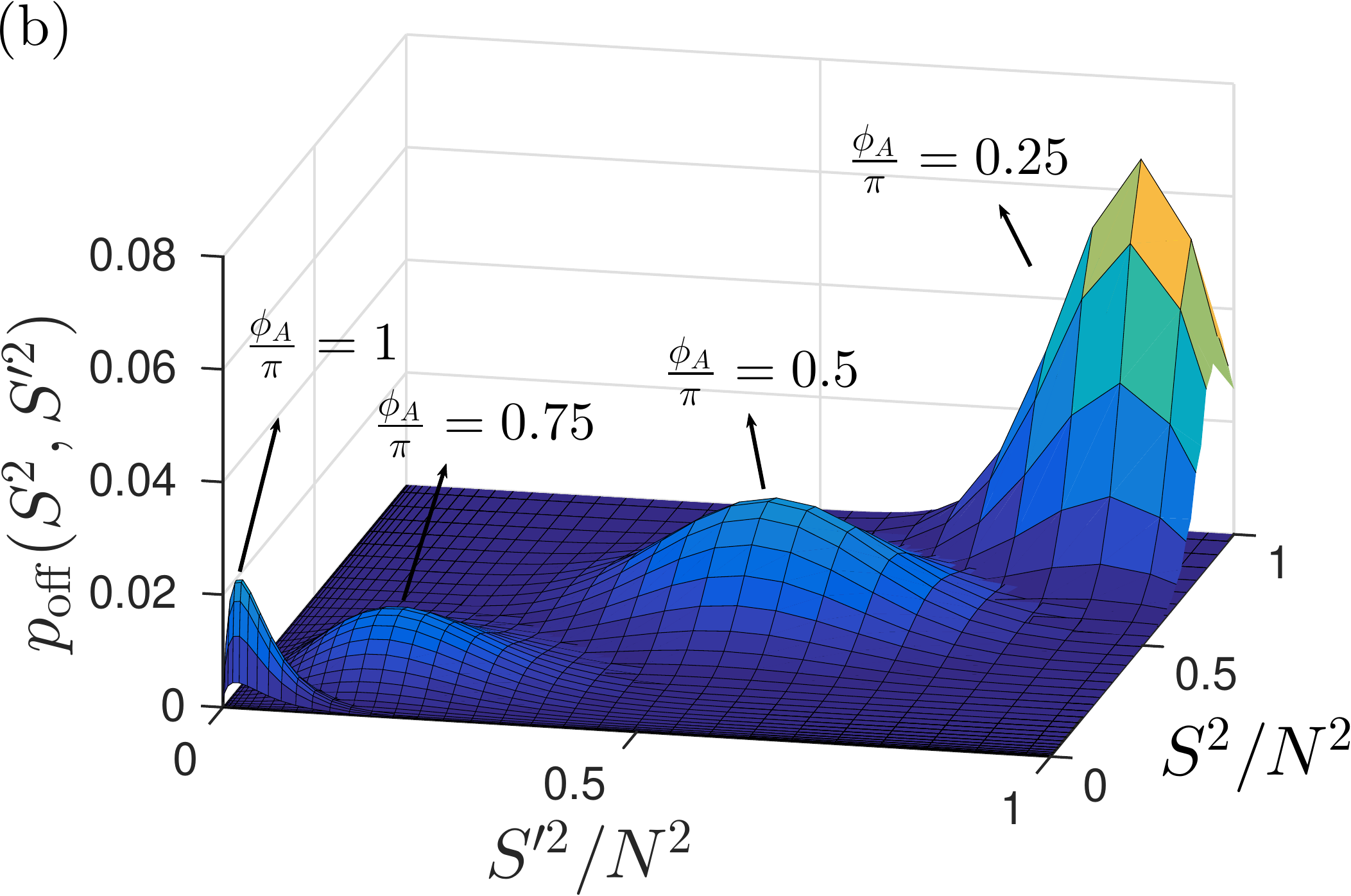}
\includegraphics[width = 0.32 \linewidth]{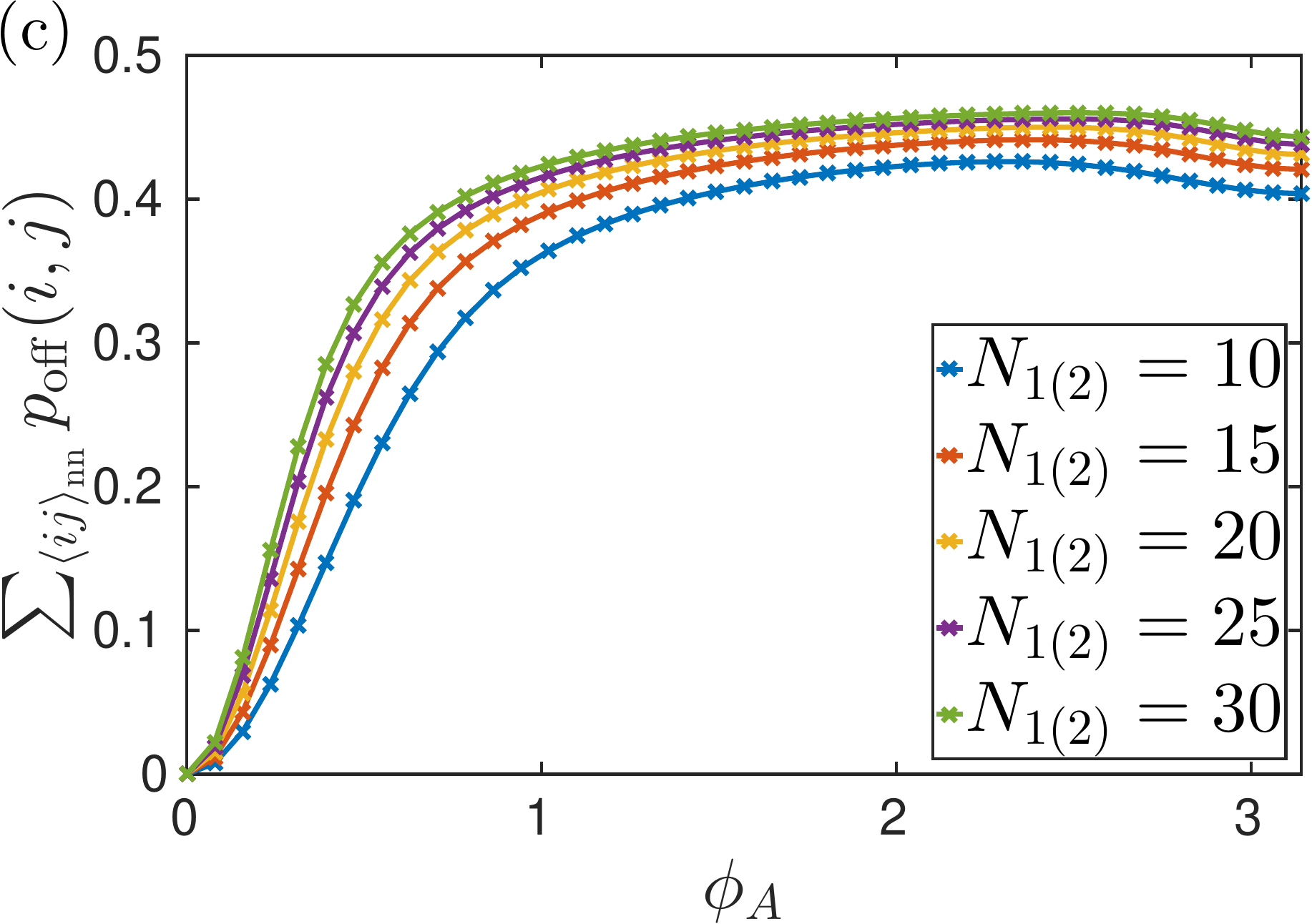}
 \caption{ Distribution of the initial state along the different angular angular subspaces for a system equivalent to $M=2$ subensembles, each with an equal number of spins $N_{1}=N_2$. We set $\theta_{1(2)} = \pi/2$, $\phi_1 = 0$ in all figures. (a) We show the diagonal distributions, for varying system sizes and inhomogeneity $\phi_A$. (b) We show the full elements, for different initial states and fixed $N_{1(2)}=32$ spins. (c) We show the integrated nearest neighbor off-diagonal elements for different system sizes and anisotropies.
 }
 \label{fig.initial.state.dependence}
\end{figure*}

We consider in this section a simple example in order to grasp the main effects of inhomogeneity in the initial preparation of the system. We assume the ensemble of $N$ spins break the permutationally invariance among a macroscopic number of its spins constituents. Breaking the symmetry at a finite number $\mathcal{O}(1)$ of spins should lead   to negligible effects in the thermodynamic limit ($N \rightarrow \infty$). We therefore consider the case where the spin ensemble is splitted into $M=2$ sub-ensembles, each being permutationally invariant among its internal spins but with the two macrosocpic ensembles breaking the symmetry among them -  as illustrated in Fig.\eqref{fig.M2.ensembles}. The system is 
equivalent to a pair of $M$ spin ensembles with quantum numbers $\{S_j^2,S_{z,j}\}_{j=1}^M$ and corresponding $N_j$ spins, evolving with isotropic collective couplings in the Lindbladian and  importantly 
conserving the relative angle between their total angular momentum, $\Re( \vec{S}_1 . \vec{S}_2)$. This latter relation arises directly from the conservation of the full system total angular momentum $S^2$ once translated into the subensembles. It induces an intrinsic correlation over the subensembles and  is a crucial aspect for their consequent peculiar dynamics, as we discuss later.

\subsection*{Algebra of inhomogeneous observables}

Despite the inhomogeneity of the system, capturing its effects depend on the observables under scrutiny. 
From one side, symmetric collective observables $\hat  O_{\mathcal{S},\alpha} = \hat S_{\alpha,1} + \hat S_{\alpha,2}$ satisfy the usual $SU(2)$ algebra and therefore can capture the dynamics only  within the diagonal subspaces of the density matrix. On the other side, in order to capture the coherences (off-diagonal terms) arising from the inhomogeneity of the initial state, one should aim for inhomogeneous observables as well, as e.g. antisymmetric observables among the two spins ensembles, $ \hat O_{\mathcal{A},\alpha} = \hat S_{\alpha,1} - \hat S_{\alpha,2}$.
Similarly to the total angular momentum which can be written as a symmetric decomposition $S^2 = \sum_{\alpha} \hat O_{\mathcal{S},\alpha}^2$, one can also define an antisymmetric total angular momentum as $S_{\mathcal{A}}^2 = \sum_{\alpha} \hat O_{\mathcal{A},\alpha}^2 $, which is conserved in the dynamics and interpreted as the total spin along the inhomogeneous subspace.

In order to unravel how the antisymmetric  observables are related to coherence terms we perform a combination of analytical and numerical studies. From the commutation relations of the subensemble spin operators we obtain that,
\begin{eqnarray}
 \hat  O_{\mathcal{S},z} (\hat  O_{\mathcal{A}}^\pm |S^2,S_z\rangle ) &=& (\hat  O_{\mathcal{A}}^\pm \hat  O_{\mathcal{S},z} + [\hat  O_{\mathcal{S},z}, \hat  O_{\mathcal{A}}^\pm ]) |S^2,S_z\rangle \nonumber \\
 &=& (S_z \pm 1 ) \hat  O_{\mathcal{A}}^\pm |S^2,S_z\rangle
\end{eqnarray}
thus relating only $S_z \rightarrow S_z \pm 1$ quantum numbers, where $\hat  O_{\mathcal{A}}^\pm = \hat  O_{\mathcal{A},x} \pm i \hat  O_{\mathcal{A},y}$ and in the second line we used that $[\hat  O_{\mathcal{S},z}, \hat  O_{\mathcal{A}}^\pm ] = \pm  \hat  O_{\mathcal{A}}^\pm$. Moreover, using numerical exact diagonalization for a finite system size we see they connect $S^2 \rightarrow S^2 \pm 1$ quantum numbers - Fig.\eqref{fig.antiss.obs.distrib}. Therefore these observables shall capture those coherences arising from nearest neighbor angular momentum subspaces in the density matrix.

\subsection*{Role of initial state preparation}

The dynamics of the system is strongly influenced by its initial state and corresponding distribution over the quantum number and conserved quantities. We consider the case where the two spin ensembles are initially uncorrelated states  
\begin{equation}\label{eq.initial.state.1}
|\psi(0)\rangle = |\psi_1(\phi_1,\theta_1)\rangle \otimes |\psi_2(\phi_2,\theta_2) \rangle,
\end{equation} 
where  $|\psi_i(\phi_i,\theta_i)\rangle$ are spin coherent states,  
\begin{equation}\label{eq.initial.state.2}
 |\psi_i\rangle = \left( \cos(\theta_i/2) |0\rangle + e^{-i \phi_i} \sin(\theta_i/2)|1\rangle \right)^{\otimes N/2}
\end{equation}
with $0 \leq \theta_i \leq \pi$ and $0 \leq \phi_i \leq 2\pi$ the azimuthal and polar angles of the spins, respectively.

We show in Fig.\eqref{fig.initial.state.dependence} the distribution of the initial state over the quantum numbers and its dependence with the inhomogeneity between the sub-ensembles. We set initial states with equal $\theta_1= \theta_2$, and varying inhomogeneity according to their polar angular difference, $\phi_A = \phi_2-\phi_1$. In order to analyse their distributions we compute their overlap  onto the corresponding subspaces, given by
\begin{eqnarray}
p_{\rm{d}}(S^2) &=& Tr(\hat \rho_{S^2,S^2})\\
p_{\rm{off}}(S^2,S'^2) &=& Tr(\hat \rho_{S^2,S'^2}^\dagger
\hat \rho_{S^2,S'^2}),
\end{eqnarray}
for the diagonal and off-diagonal terms, respectively.

In Fig.(\ref{fig.initial.state.dependence}a) we show the distribution for the diagonal terms. We see that increasing the inhomogeneity tends to decrease the average angular momentum, therefore inducing the system towards dark subspaces with lower effective number of spins in the dynamics. For the extreme cases with $\phi_A$ close to zero (symmetric) or $\pi$ (fully assymetric) the distributions have sharper peaks, while for intermediate angles they display broader variances. It is important to notice that the average values scale extensively with the number of spins (the overlap for a fixed $S^2$ subspace of order $\mathcal{O}(1)$ vanishes in the thermodynamic limit - not shown), therefore the effective number of spins becomes macroscopic (extensive) in the    
thermodynamic limit regardless of their anisotropies.

The distribution over the off-diagonal terms follows a similar trend - see Fig.(\ref{fig.initial.state.dependence}b) - where increasing the inhomogeneity leads to broader distributions along the different subspaces. We expect stronger effects over the inhomogeneous observables for larger angles $\phi_A$. In particular, the coherences between nearest neighbor subspaces increases with $\phi_A$, as shown in Fig.(\ref{fig.initial.state.dependence}c), implying stronger impact over antisymmetric observables.

\section{The Dicke model}
\label{sec.Dicke.model}

\begin{figure*}
\includegraphics[width = 0.32 \linewidth]{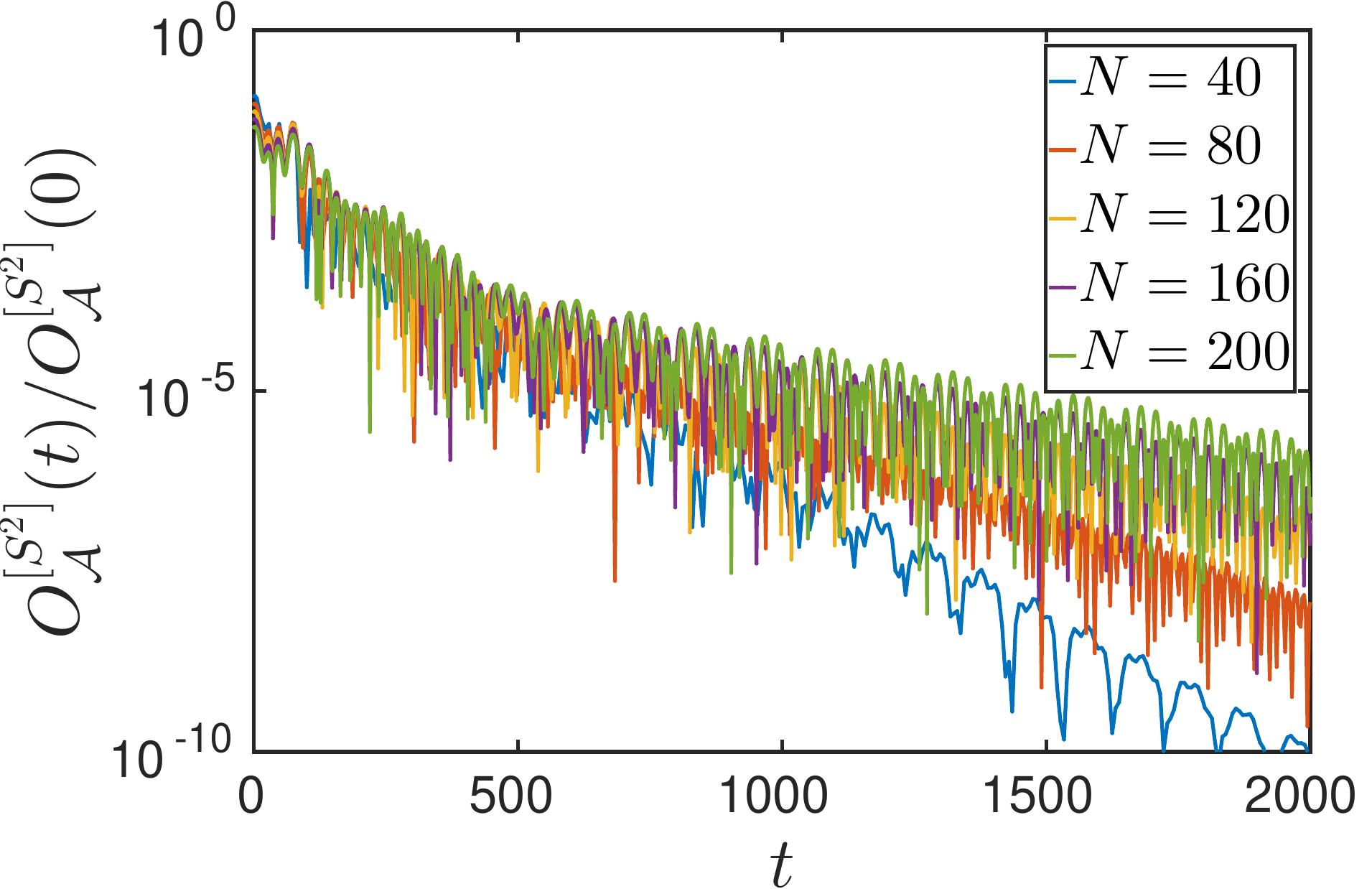}
\includegraphics[width = 0.32 \linewidth]{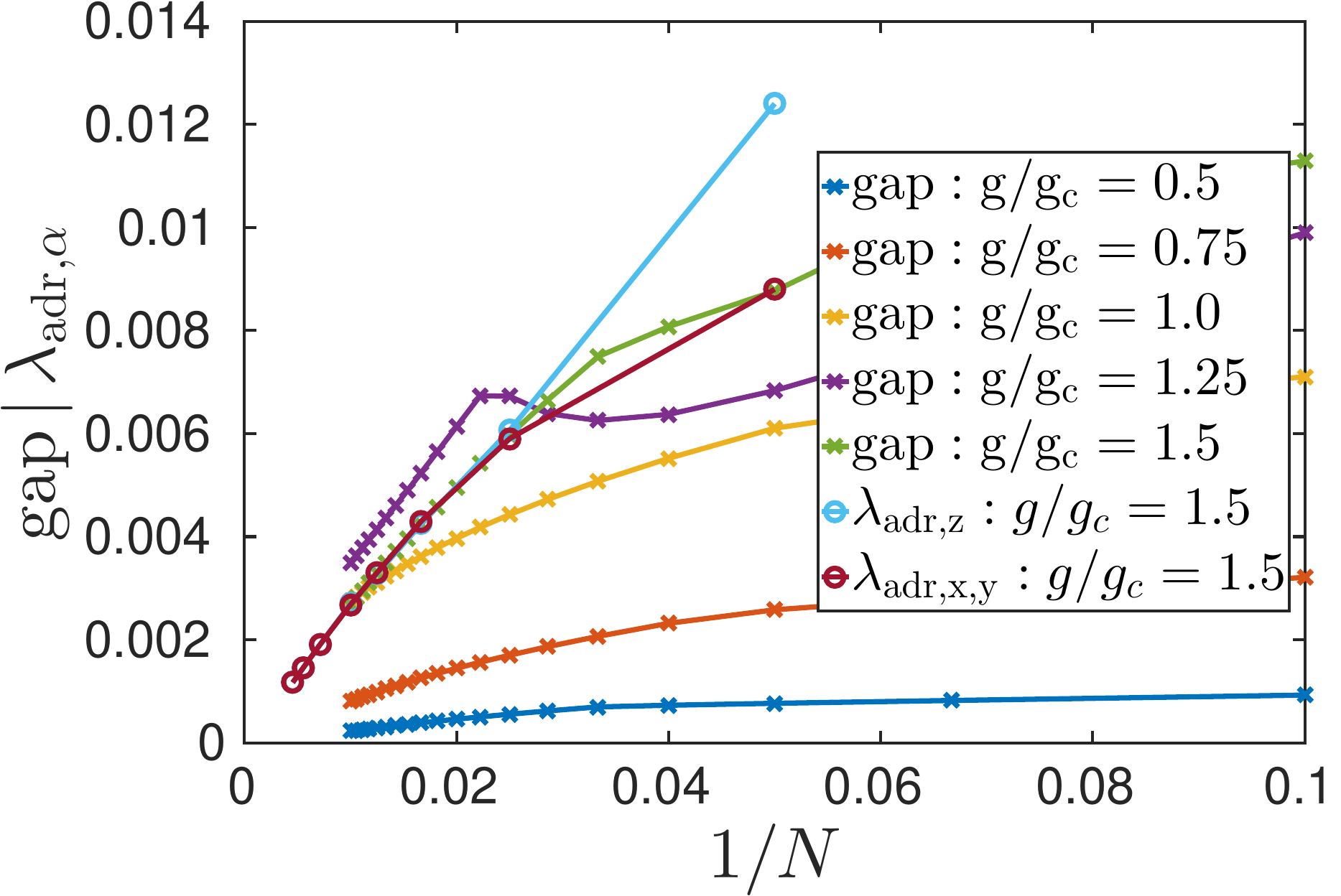}
\includegraphics[width = 0.32 \linewidth]{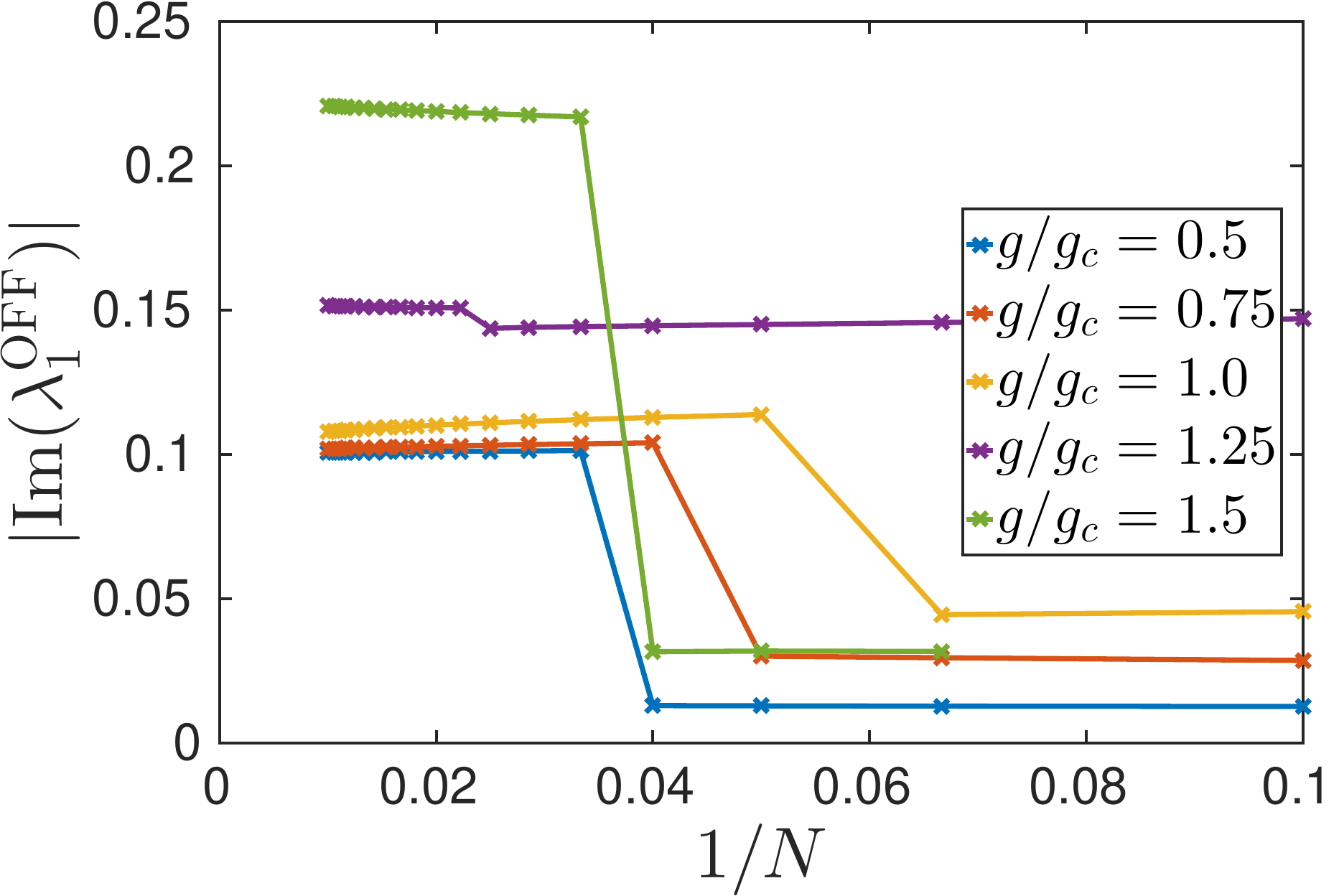}
 \caption{  
 Dynamics for the off-diagonal subspace $\hat \rho_{S^2,S^2-1}(t)$, with $S^2=N/2$, in the effective Dicke model of Eq.\eqref{eq.spinonly.Dicke} with $\omega_z=0.1$, $\omega_0=\kappa=1$. The initial inhomogeneity between the $M=2$ subensembles is setlled to $\theta_1=\theta_2$, $\phi_1=0$ and $\phi_2=0.1\pi$. 
 (a) We show the dynamics for the antisymmetric observable onto such subspace $O_{\mathcal{A}}^{[S^2]}(t) = Tr(\hat O_{\mathcal{A},x}\hat \rho_{S^2,S^2-1}(t))$, for a coupling $g=1.5g_{\rm{cr}}$ and varying number of spins. The dynamics is shown in log-linear scale highlighting the slower vanishing in time according to the gapless modes.
 (b) Lindbladian gap on the off-diagonal subspace for varying couplings $g$. We also show the assymptotic decay rate 
 $\lambda_{\rm{adr},\alpha}$ for  $g/g_{\rm{cr}}=1.5$. (c) The imaginary part of the gap eigenvalues.
 }
 \label{fig.off.diag.dynamics}
\end{figure*}

In this section we revisit the paradigmatic Dicke model within this new perspective, exploring the inhomogeneous effects over its observable dynamics. 
The full model is defined by an ensemble with $N$ spins collectively interacting with a bosonic mode \cite{Kirton_2019}, as follows,
\begin{equation}
 \frac{ d\hat \rho}{dt} = -i[\hat H, \hat \rho] + \kappa \left( \hat a \hat \rho \hat a^\dagger - \frac{1}{2}\{ \hat a^\dagger \hat a,\hat \rho \}
 \right)
\end{equation}
where $\kappa$ is the bosonic decay rate, and the coherent  Hamiltonian given by,
\begin{equation}
 \hat H = \omega_z   \hat S^z +  \omega_0 \hat a^\dagger \hat a + \frac{g}{\sqrt{N}} \hat  S^x (\hat a+\hat a^\dagger),
\end{equation}
with $\hat a (\hat a^\dagger)$ the creation (anihillation) bosonic operator satisfying commutation relations $[\hat a,\hat a^\dagger] = 1$, $\omega_z $ is a Zeeman field on the spins, $\omega_0$ the energy of the bosonic mode and $g$ the spin-boson coupling strength. The model has a $\mathbb{Z}_2$ symmetry under the transformation $\hat S^x \rightarrow -\hat S_x$,
$\hat a \rightarrow -\hat a$. Given 
 given initial states in the symmetric subsapce, i.e., with maximun angular momentum $S^2=N/2$, the steady state of the system in the thermodynamic limit (within a mean-field approach) displays a phase transition between a normal phase (NR) to a superradiant phase (SR). While the normal phase is characterized by $\langle  \hat S^x \rangle = \langle \hat a + \hat a^\dagger \rangle  = 0$, the superradiant phase spontaneously break the symmetry with $\langle \hat S^x \rangle \neq 0$. The critical coupling for the transition occurs at $g_{\rm cr}^2 = \omega_z (\omega_0^2 + \kappa^2)/(\omega_0)$.

\subsection*{Spin-only dynamics}

The analysis of the quantum dynamics with finite $N$ spins in the full model is usually demanding, either within numerical or analytical approaches. We therefore study the dynamics within an effective spin-only Lindbladian following Ref.\cite{Jager_2022}, obtained by tracing out the bosonic degrees of freedom concomitant to a Schrieffer-Wolff transformation. The effective Lindbladian is in the form of Eq.\eqref{eq.spin.gen.Lindbladian},
as follows,
\begin{equation}\label{eq.spinonly.Dicke}
 \mathcal{L}[\hat \rho] = -i[\hat H_{\rm{eff}},\hat \rho] + 
 \kappa \left[ \hat D \hat \rho \hat D^\dagger - \frac{1}{2}\{ \hat D^\dagger \hat D,\hat \rho \} \right] 
\end{equation}
with the second term corresponding to the effective dissipation, and the first one given by the following Hamiltonian,
\begin{equation}
 \hat H_{\rm{eff}} = \omega_z \hat S^z + \frac{g}{2\sqrt{N}}(\hat S^x \hat D + \hat D^\dagger \hat S^x)
\end{equation}
The displacement operator $\hat D$ includes the fluctuations and boson mediated interactions in the spin system,
\begin{equation}
  \hat D = \alpha_+ \hat S_+ + \alpha_- \hat S_-,
 \end{equation}
 with $\alpha_{\pm} = -g/[2\sqrt{N} (\omega_o \pm \omega_z -i\kappa)]$.

We study the dynamics from the initial states as defined in Eqs.(\ref{eq.initial.state.1})-(\ref{eq.initial.state.2}), with $M=2$ inhomogeneous spin subensembles and each with the same number of spins, $N_{1(2)}=N/2$. The effects on the diagonal terms of the full Dicke basis decomposition is to shift the spin-boson coupling due to their different effective number of spins $N_{\rm eff} = 2S^2$ for the different subspaces. Therefore larger anisotropies lead to  stronger couplings for the superradiant transition in the model.

We find an interesting, and unexpected, dynamics for the off-diagonal terms. From one side, due to the decoherent dissipative nature of the dynamics, the coherences as expected vanishes in the long time limit, i.e., with off-diagonal $\hat \rho_{S^2,S'^2}(t) \rightarrow 0$, for $t \rightarrow \infty$. We show in Fig.(\ref{fig.off.diag.dynamics}a) the dynamics for nn off-diagonal subspaces $\hat \rho_{S^2,S^2-1}(t)$ for an inhomogeneous initial condition. On the other side, we find that the vanishing dynamics is slower as one increases the system size. 
The dynamics at long time evolves exponentially in time towards its steady state value, where $O_{\mathcal{A},\alpha} \sim e^{-\lambda_{\rm adr,\alpha}t}$ with $\lambda_{\rm adr,\alpha}$ the asymptotic decay rate.
The long time dynamics can also be captured by the Lindblad gap of the constrained subspace, i.e., 
\begin{equation}
 \rm{gap}(S^2,S'^2) = -\max_{\{ \lambda \, || \, \Re(\lambda)\neq 0 \}} \Re (\lambda_{S^2,S'^2})
\end{equation}
with $\lambda_{S^2,S'^2}$ the eigenvalues of the projected Lindbladian onto the corresponding off-diagonal subspaces, $\hat P_{S^2} \mathcal{L} \hat P_{S'^2}$.
 Performing a finite-size scaling of the asymptotic decay rate and to the Lindbladian gap 
we obtain they both correspond to gapless excitations following a $1/N$ scaling, as shown in Fig.(\ref{fig.off.diag.dynamics}b). Moreover, we notice that the gapless eigenvalues have non-vanishing imaginary terms - Fig.(\ref{fig.off.diag.dynamics}c) - implying nontrivial dynamics within their corresponding eigenspaces \cite{Leonardo2023}. Therefore, the system in the thermodynamic limit can preserve its coherences among different angular momentum subspaces, possibly  leading to nontrivial off-diagonal steady states or emergent persistent dynamics.

\subsection*{Mean-field}

\begin{figure*}
\includegraphics[width = 0.34 \linewidth]{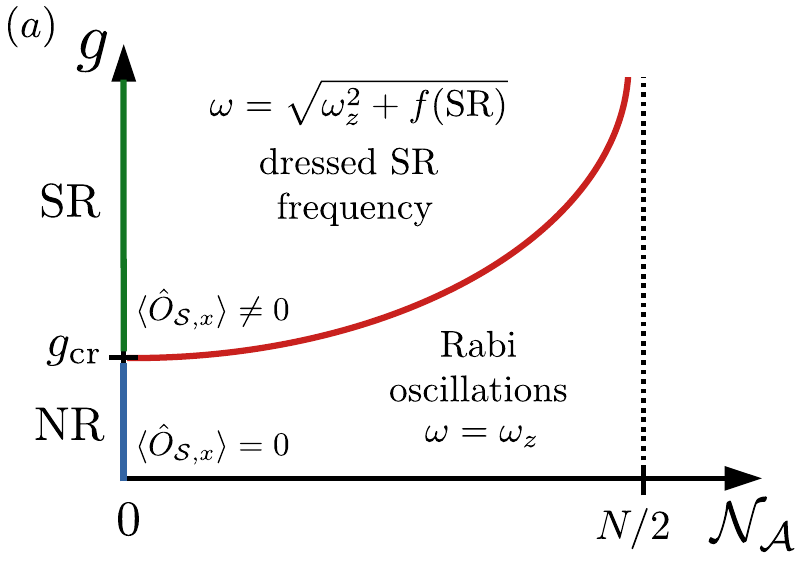}
\includegraphics[width = 0.32 \linewidth]{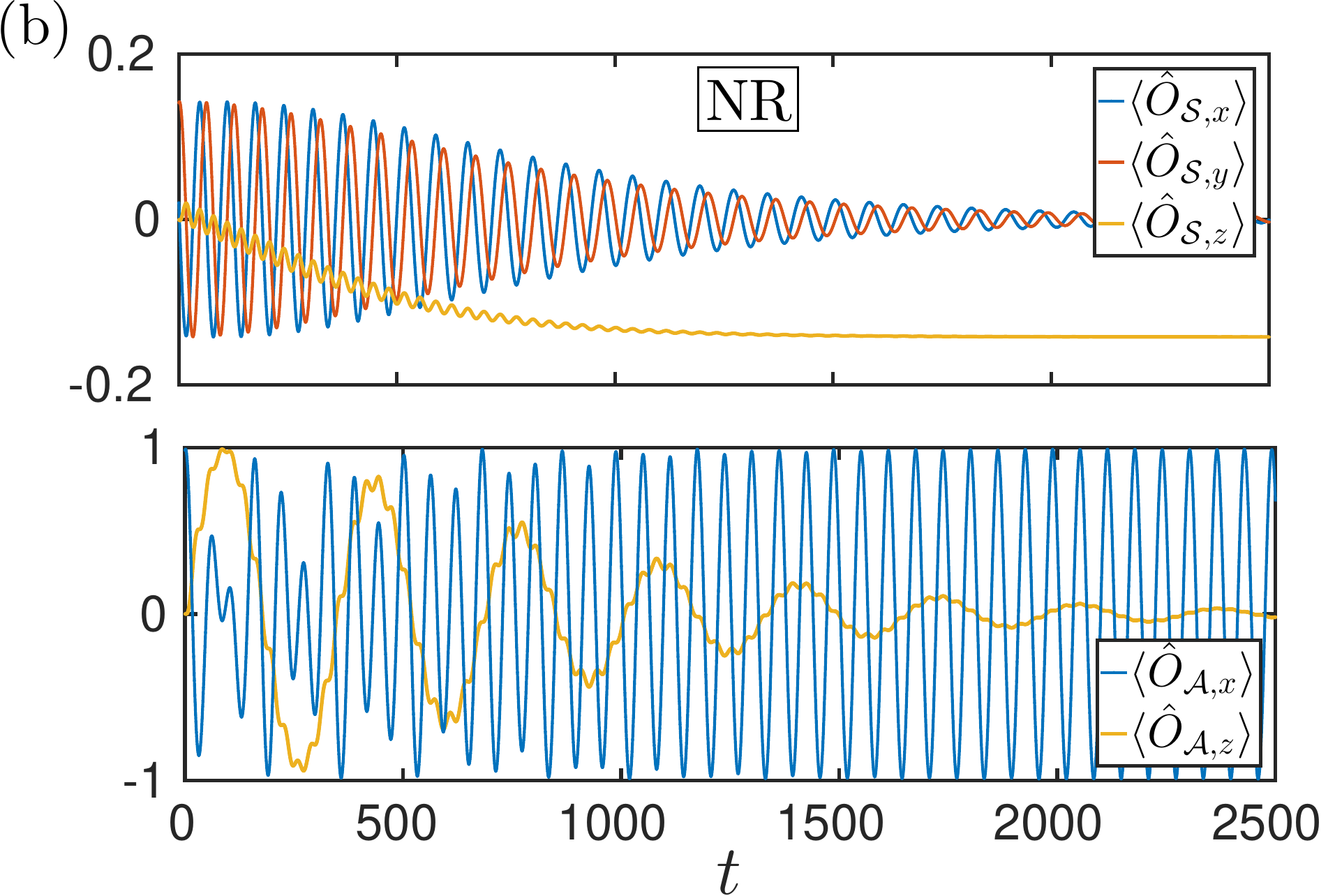}
\includegraphics[width = 0.32 \linewidth]{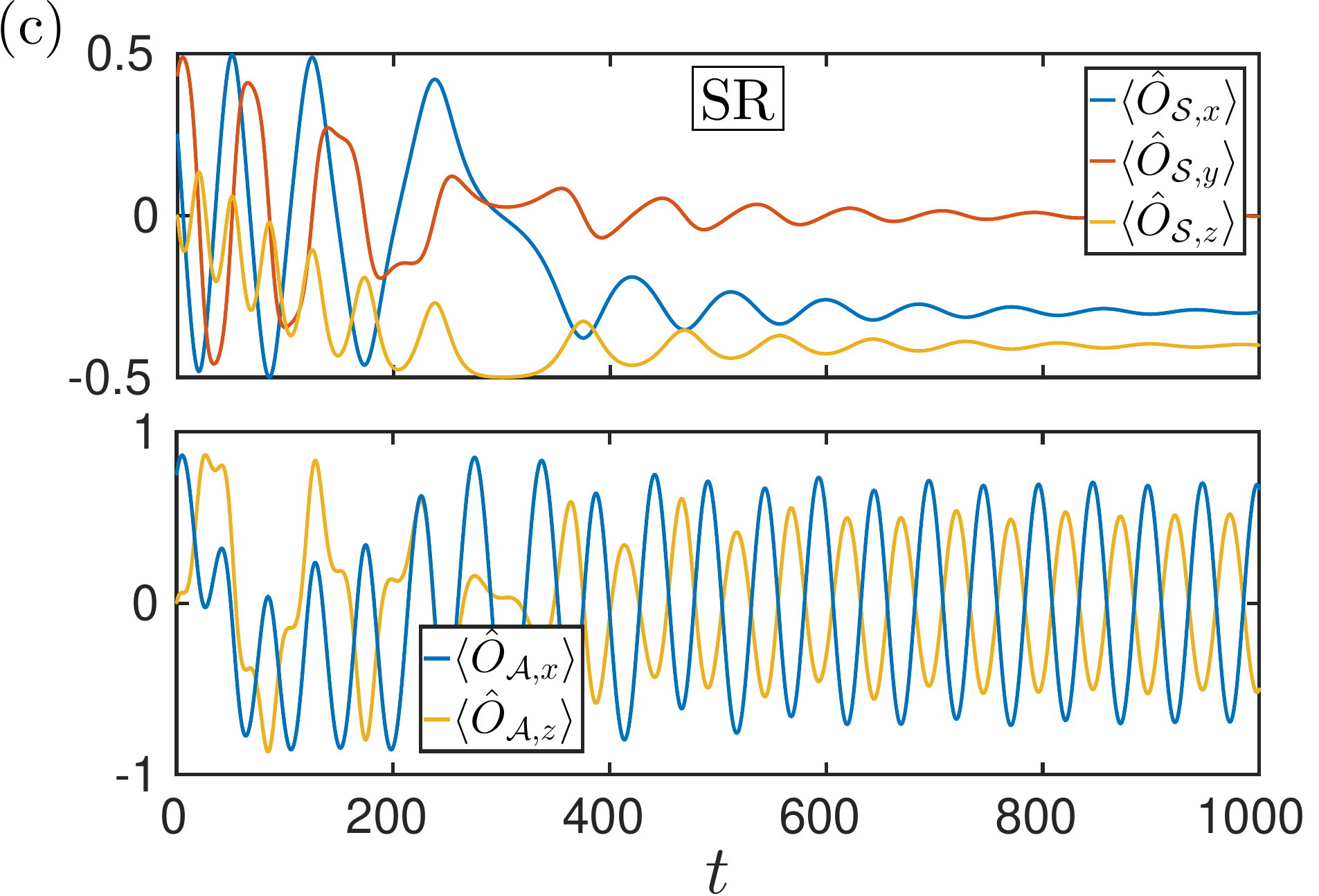}
 \caption{  (a) Mean-field phase diagram for the Dicke model with inhomogeneous initial states, equivalent to $M=2$ subensembles. The critical coupling $g$ depends on the initial inhomogeneity $\mathcal{N}_{\mathcal{A}}$ between the spin ensembles. For inhomogeneous states we see the emergence of limit cycles with (b) Rabi frequencies or (c) dressed by the SR magnetization, both captured by antisymmetric observables. In both plots we used $\omega_z=0.1$, $\omega_o=\kappa=1$, $g=0.5$, $\theta_{1(2)}=0$ with $\phi_A = \phi_1-\phi_2=\pi/1.5$ and  $\phi_A = \pi/1.1$ for the NR and SR phases, respectively. The shifted frequency is given by  $\rm{f(SR)} =
 [ J_x( O_{\mathcal{S},x})_{\rm{ss}} + J_y( O_{\mathcal{S}, y})_{\rm{ss}} ]^2 $.
 }
 \label{fig.MF.phase.diagram}
\end{figure*}

In order to study these peculiar effects  in the thermodynamic limit we employ a mean-field (MF) analysis for the spins, which is exact on such all-to-all interacting systems \cite{Bhaseen2012}. The mean-field approach closes the expectation values at the second order cumulant, with factorizable expectation values $\langle \hat S^\alpha_i \hat S^\beta_j \rangle = \langle \hat S^\alpha_i  \rangle \langle \hat S^\beta_j \rangle$. The equations of motion for the observables become in this way closed within MF. We focus our studies on the symmetric observables ($\hat O_{\mathcal{S},\alpha}$) and antisymmetric ones ($\hat O_{\mathcal{A},\alpha}$), with EOM's given by
 \begin{eqnarray}
 \overset{.}{O}_{\mathcal{S},x} &=& -\omega_z {O}_{\mathcal{S},y}, \label{eom.S.x}\nonumber  \\
  \overset{.}{O}_{\mathcal{S},y} &=&\omega_z {O}_{\mathcal{S},x} -{O}_{\mathcal{S},z} \left(   J_x {O}_{\mathcal{S},x} + J_y {O}_{\mathcal{S},y} \right), \label{eom.S.y} \\
  \overset{.}{O}_{\mathcal{S},z} &=&  {O}_{\mathcal{S},y} \left(    J_x {O}_{\mathcal{S},x} + J_y  {O}_{\mathcal{S},y}\right), \label{eom.S.z}\nonumber 
 \end{eqnarray}
 and 
 \begin{eqnarray}
  \overset{.}{O}_{\mathcal{A},x} &=& -\omega_z {O}_{\mathcal{A},y}, \label{eom.A.x} \nonumber \\
  \overset{.}{O}_{\mathcal{A},y} &=&\omega_z {O}_{\mathcal{A},x} - {O}_{\mathcal{A},z} \left(   J_x {O}_{\mathcal{S},x} +  J_y {O}_{\mathcal{S},y}\right), \label{eom.A.y} \\
  \overset{.}{O}_{\mathcal{A},z} &=&  {O}_{\mathcal{A},y} \left(  J_x {O}_{\mathcal{S},x} + J_y  {O}_{\mathcal{S},y}  \right), \label{eom.A.z} \nonumber 
 \end{eqnarray}
respectively, where $J_x = 2g\Re(\alpha_+ + \alpha_-)/\sqrt{N}$ and $J_y= -2 g\Im(\alpha_+ . \alpha_-)/\sqrt{N}$.   
The  phase diagram of the system is shown in Fig.(\ref{fig.MF.phase.diagram}a), displaying  the appearance of inhomogeneity-dependent critical coupling for the SR-NR phase transition, as well as the emergence of LC dynamics with frequencies driven by Rabi oscillations or dressed by SR magnetization.

In order to derive the phase diagram we first notice that EOM's have emergent conserved quantities given by the norm $\mathcal{N}_{\mathcal{S}/\mathcal{A}}= \sum_{\alpha=x,y,z} (O_{\mathcal{S},\mathcal{A}}^\alpha)^2$ onto the symmetric/antisymmetric MF observables,  with $d \mathcal{N}_{\mathcal{S},\mathcal{A}}/dt = 0$. Therefore the steady states and dynamics are strongly influenced by the initial conditions and their corresponding norms onto such subspaces. 

We see that the symmetric observables evolves as a closed set, i.e.,  depending only on symmetric observables as well. In particular, their evolution is, as expected, similar to the usual Dicke model approach but with an effective number of spins, leading to the shifted critical coupling between SR and NR phases,
\begin{equation}
 g_{\rm cr}^2 = \frac{\omega_z (\omega_0^2 + \kappa^2)}{( \omega_0)} \left(\frac{N}{2\mathcal{N}_{\mathcal{S}}} \right)
\end{equation} 
The antisymmetric terms on the other side depend not only on themselves, but also on the symmetric ones. In this way, their long time dynamics are influenced by the the steady state observables of the diagonal subspaces. Assuming the system reaches its steady state magnetization $\{ (O_{\mathcal{S}, \alpha})_{\rm{ss}} \}_{\alpha=x,y,z}$, we can obtain the second derivative equation of motion for the antisymmetric observables,
 \begin{equation}\label{eq.harmonic.osc.antiss}
  \overset{..}{O}_{\mathcal{A},y}  = -\left[\omega_z^2 + [ J_x( O_{\mathcal{S},x})_{\rm{ss}} + J_y( O_{\mathcal{S}, y})_{\rm{ss}} ]^2 \right] {O}_{\mathcal{A},y} 
 \end{equation}
following an harmonic oscillator dynamics, whose frequency depend on the underlying NR (SR) phase of the system.  Specifically, the frequency of the oscillations are given by the Rabi oscillation when the diagonal steady state is in the NR phase, while it is nontrivially shifted (dressed) in the SR phase. The amplitude of the oscillation is given by square root of the norm ($\sqrt{\mathcal{N}_\mathcal{A}}$) onto the subspace.

\subsection*{General inhomogeneity}

We expand here the MF discussion for general $M$'th order anisotropies. Specifically, we consider the case where the ensemble can be splitted into $M$ subensembles with spins $\{ S_{\alpha,i} \}_{i=1}^M$, each containing $N_i$ spins which are permutationally invariant among each other, but with the subensembles breaking globally the symmetry.

In order to generalize the analysis, we first introduce the spin vector notation $\vec{S}_i = (S_{x,i},S_{y,i},S_{z,i})$, for which the EOM's are described in the simpler torque vector formalism. Specifically, the EOM's are given as follows,
\begin{eqnarray}
  d\vec{S}_i/dt = \vec{\tau} \times \vec{S}_i,
 \end{eqnarray}
 where $\vec{\tau} = \vec{\tau}_{\omega_z} +  \vec{\tau}_{J_x} + \vec{\tau}_{J_y}$ is the total torque vector acting on the $i$'th spin subensemble. Their components are given by,
 \begin{eqnarray}
  \vec{\tau}_{\omega_z} &=& (0,0,\omega_z), \nonumber \\
  \vec{\tau}_{J_x} &=& (J_x O_{\mathcal{S},x},0,0), \\ \,
  \vec{\tau}_{J_y} &=& (J_y O_{\mathcal{S},y},0,0),\nonumber
 \end{eqnarray}
 From this formalism we directly notice that spin configurations with vanishing   mean magnetization  are \textit{dark} to  dynamics, i.e., they cancel the net effects of the interactions among the subensembles, which are now driven only by the external Zeeman fields applied to each of them. Although these configurations may still not be steady states of the evolution, they provide a route to decompose the EOM's of the system.

 The dark configurations correspond to the zero eigenvalues of the all-ones $M \times M$ matrix, defined as $(\hat 1)_{i,j=1}^M=1$ and zero otherwise. The all-ones matrix has a single non-vanishing eigenvalue for the totally symmetric vector,
 \begin{equation}
  \lambda_{\mathcal{S}}^{[\hat 1]} = M, \qquad \vec{v}_{\mathcal{S}} = (1,1,...,1)^T/\sqrt{M}
 \end{equation}
 with all other $M-1$ eigenvalues null, denoted by $\{\lambda_{k}\}_{i=1}^{M-1}$ with corresponding 
 $\vec{v}_k$ eigenvectors. The eigenvector for the null subspace in the case of $M=2$ is simply the antisymmetric vector $(1 ,-1)/\sqrt{2}$ while for $M>2$ it has slightly more complicated forms (neverthless with analytical form).

\begin{figure}
\includegraphics[width = 0.8 \linewidth]{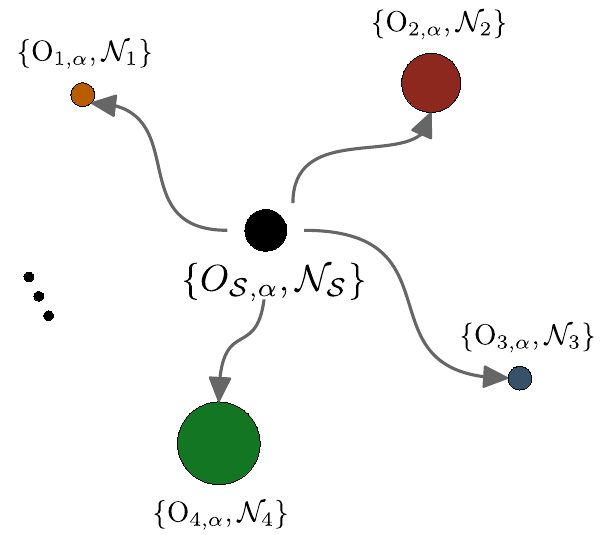}
 \caption{   Schematic representation for the equations of motion of the inhomogeneous modulated observables $\{ O_{k,\alpha} \}$, each with norm $\mathcal{N_k}$ depending on the initial state. All of them are correlated by their own elements and to the symmetric dynamics $\{ O_{\mathcal{S},\alpha} \}$. The reverse is not true, with the symmetric observables constituting a closed set (unidirectionality in the EOM's).
 }
 \label{fig.gen.MF.inhomogeneity}
\end{figure}

 It is now convenient to study the global observables for the spin subensembles based on the dark and symmetric eigenstates. We define the global observables modulated by these eigenvectors,
 \begin{equation}
  O_{k,\alpha} = \sum_{i=1}^M (\vec{v}_k)_i S_{\alpha,i}, 
 \end{equation}
 and $\hat O_\mathcal{S,\alpha} $ obtained from the symmetric vector $\vec{v}_{\mathcal{S}}$ - recovering the definition as in the previous sections. The EOM's now decouple in a specific structure, 
 \begin{eqnarray}
  \overset{.}{O}_{\mathcal{S},\alpha} &=& f_\mathcal{S}(\{
   O_{\mathcal{S},\alpha'} \}), \nonumber \\
   \overset{.}{O}_{k,\alpha} &=& f_k(\{
   O_{\mathcal{S},\alpha'}, O_{k,\alpha'} \}),  
 \end{eqnarray}
where $f_{\mathcal{S}(k)}$ are nonlinear functions of their arguments. We see that while the symmetric observables are a closed set, the inhomogeneous observables in such a basis depend on their own elements and the symmetric ones - as schematically represented in Fig.\eqref{fig.gen.MF.inhomogeneity} -  generalizing the results for $M=2$. Moreover, each subset with fixed $k$ conserves its norm, therefore we can also expect persistent oscillations or LC dynamics of the corresponding observables.

 \section{Boundary Time Crystal}
 \label{sec.btc}

 In this section we revisit the dynamics of a boundary time crystal phase in light of the theory developed in the previous sections. 
 Boundary time crystals are nonequilibrium phases of matter occurring in quantum systems in contact
to an environment, for which a macroscopic fraction of the many-body system displays robust limit cycles \cite{Iemini2018_btc}. A paradigmatic model manisfesting this behaviour is given by the Lindbladian,
\begin{equation}\label{eq.Lindbladian.btc}
 \mathcal{L}[\hat \rho] = -i [\hat H,\hat \rho] + \frac{2\kappa}{N} \left( \hat S_{-} \hat \rho \hat S_+ - \frac{1}{2}\{\hat S_- \hat S_+, \hat \rho\} \right)
\end{equation}
with coherent Hamiltonian,
\begin{equation}
 \hat H = \omega_x \hat S^x + \frac{2J_{xx}}{N} (\hat S^x)^2,
\end{equation}
where $\kappa$ represents the collective dissipative decay of the ensemble, $\omega_x$ is a global magnetic field  and $J_{xx}$ spin interactions along the field direction. The model in its simplest form with $J_{xx}=0$ 
(for equilibrium properties of the model see~\cite{Walls1978,Puri1979,Larson2018}) 
shows a phase transition between a ferromagnetic stable steady state with $\langle \hat S^z \rangle_{\rm{ss}} \neq 0$ to a time crystal phase with unstable steady states and persistent oscilations of its macroscopic magnetization in the thermodynamic limit. The critical coupling for the transition occurs at $\omega_x/\kappa = 1$. The BTC is robust to perturbations and the presence of interactions  $J_{xx}$ can modify its phase space portrait, with the emergence of new stable (ferromagnetic) and unstable (paramagnetic) steady states \cite{Prazeres2021}.

We consider an inhomogeneous system equivalent to $M=2$ subensembles, each with the same number of spins $N_1=N_2=N/2$. In the thermodynamic limit and within a mean-field approach, the EOM's for the symmetric and antisymmetric observables are given by,
\begin{eqnarray}
 \overset{.}{O}_{\mathcal{S},x} &=& \kappa O_{\mathcal{S},z} O_{\mathcal{S},x} \nonumber  \\
  \overset{.}{O}_{\mathcal{S},y} &=& O_{\mathcal{S},z} \left(\kappa  O_{\mathcal{S},y} - \omega_x \right) \\
  \overset{.}{O}_{\mathcal{S},z} &=&  -\kappa (O_{\mathcal{S},x}^2 + O_{\mathcal{S},y}^2) + \omega_x O_{\mathcal{S},y} \nonumber 
 \end{eqnarray}
 and 
 \begin{eqnarray}
  \overset{.}{O}_{\mathcal{A},x} &=& \kappa O_{\mathcal{A},z} O_{\mathcal{S},x} \nonumber  \\
  \overset{.}{O}_{\mathcal{A},y} &=&  O_{\mathcal{A},z} \left(\kappa  O_{\mathcal{S},y} - \omega_x \right) \\
  \overset{.}{O}_{\mathcal{A},z} &=&  -\kappa (O_{\mathcal{A},x}O_{\mathcal{S},x} + O_{\mathcal{A},y}O_{\mathcal{S},y}) + \omega_x O_{\mathcal{S},y} \nonumber 
 \end{eqnarray}
 
The  equations of motion have the structure as discussed in the previous section, with a closed symmetric sector and an antisymmetric one evolving under an ``external field'' mediated by the symmetric sector. It is interesting to notice that when the system supports a BTC phase, therefore with persistent oscillations in the symmetric sector, the external field seen by the anti-symmetric sector will be periodic and time-dependent, even for infinitely long times. Indeed, the   dimensionality for the EOM's in the antisymmetric sector effectively increases due to the time-dependent dimension, allowing more general forms of steady state fixed points and dynamical phases. 
Specifically, in the usual case of a symmetric sector with time-independent steady state,  the dimension of the $\mathcal{A}$ sector in the long-time limit is $d_{\mathcal{A}} = 3-1=2$, i.e., $3$ dimensions from the magnetization degrees of freedom and a reduction due to conservation of norm $\mathcal{N}_{\mathcal{A}}$. In this case, since the equations of motion are effective two-dimensional the possible dynamics are constrained, forbidding e.g. chaos due to the Poincaré–Bendixson theorem \cite{Strogatz_book} which imposes minimal $3$-dimensional EOM's. In the presence of BTC's, however,  the effective dimension can increase to $d_{\mathcal{A}} = 3$, allowing richer dynamical phases.

  \begin{figure}
\includegraphics[width = 0.9 \linewidth]{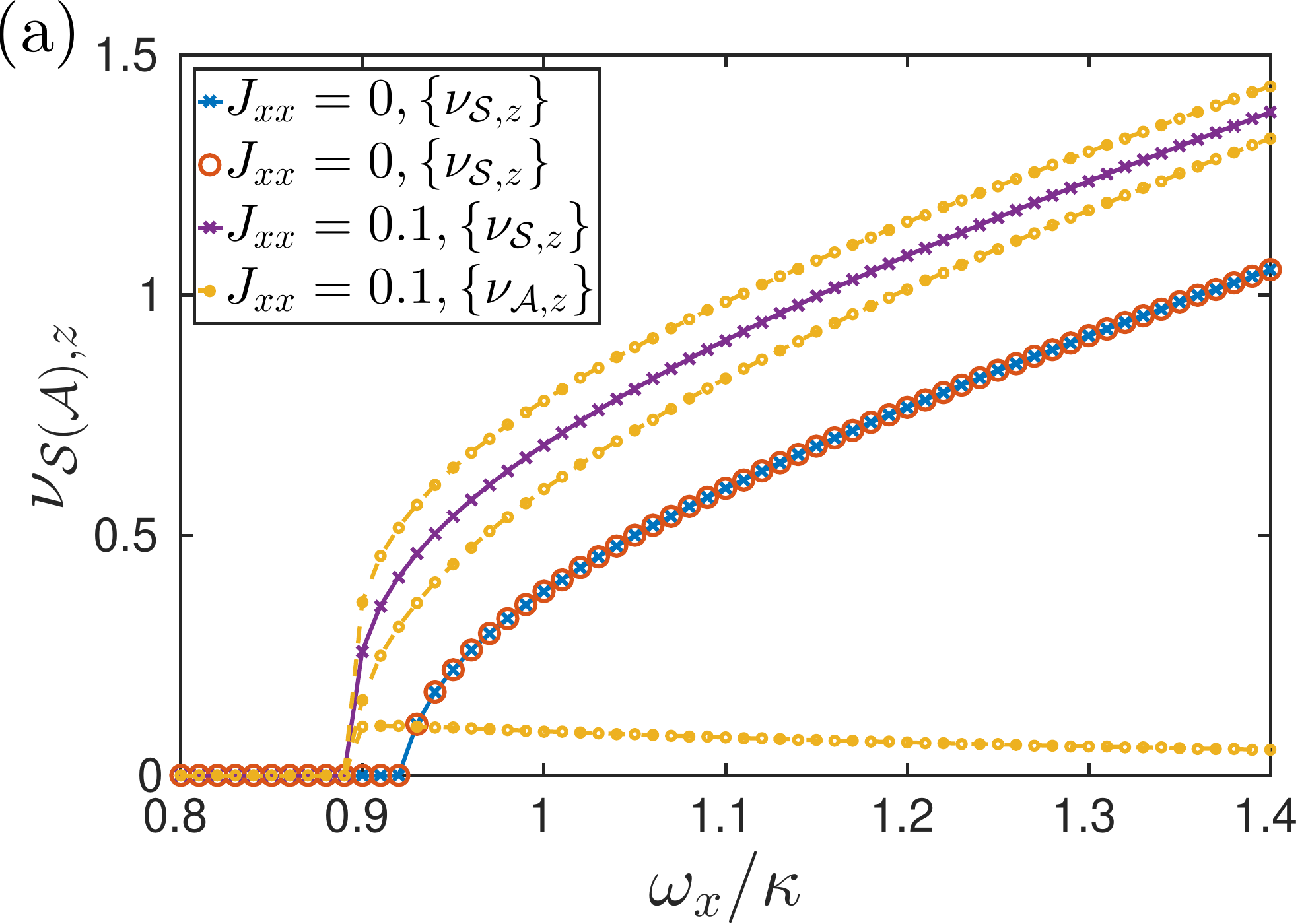}
\includegraphics[width = 0.9 \linewidth]{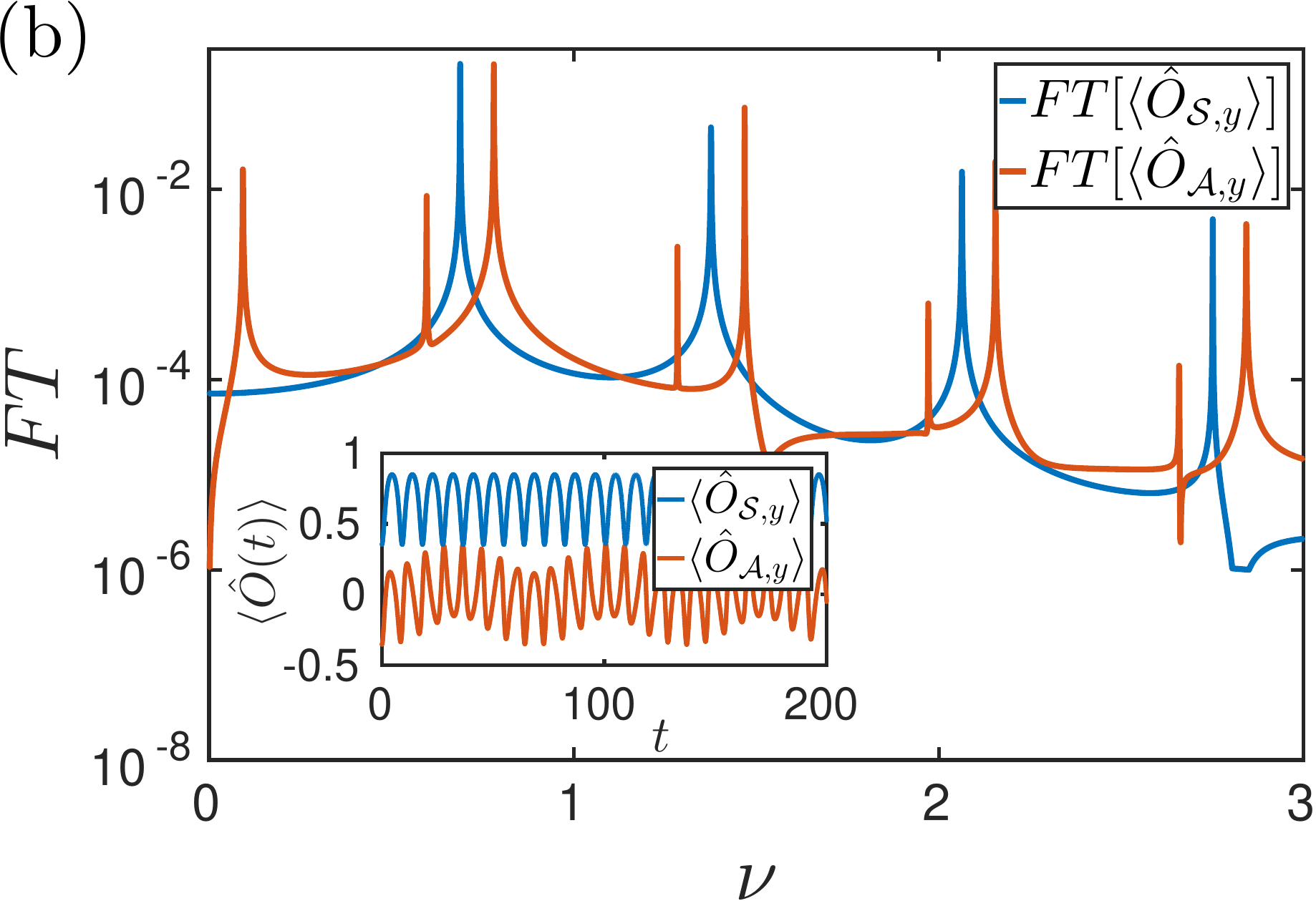}
 \caption{ Dynamics for of $M=2$ subsensembles with $N_1=N_2=N/2$, in a BTC model - Eq.\eqref{eq.Lindbladian.btc}. The initial state is set to $\theta_{1(2)}=\pi/2$ and $\phi_1=0$, $\phi_2=\pi/4$. (a) Phase diagram showing the dominant frequencies $\nu_{\mathcal{S}(\mathcal{A}),\alpha}$ for the long-time dynamics of the corresponding observable $O_{\mathcal{S}(\mathcal{A}),\alpha}$. The model display three different dynamical phases: a time-independent steady state (zero frequency), or persistent dynamics with either locked frequencies ($J_{xx} = 0$) or with beatings ($J_{xx} = 0$). We show in (b) the Fourier spectrum for the dynamics with beating frequencies, for couplings $\omega_x/\kappa = 1$ and $J_{xx} = 0.1$. The inset displays beats in the dynamics for symmetric and antisymmetric observables. 
 }
 \label{fig.btc}
\end{figure}

We show our results in Fig.\eqref{fig.btc}. In the case with $J_{xx} = 0$ we observe that the inhomogeneity enlarges the BTC phase, decreasing the critical coupling $(\omega_x/\kappa)_{\rm{cr}}$ for the phase transition. While in the ferromagnetic phase ($\omega_x/\kappa < (\omega_x/\kappa)_{\rm{cr}} $) both symmetric and antisymmetric observables stabilize to a constant value in the long-time limit, in the BTC ($\omega_x/\kappa \geq (\omega_x/\kappa)_{\rm{cr}} $) the mediated external field induces persistent dynamics onto the antisymmetric sector. Moreover, their oscillations are locked in frequency, as shown in Fig.(\ref{fig.btc}a) displaying the main frequency $\nu_{\mathcal{S}(\mathcal{A}),\alpha}$ obtained by a Fourier transformation for  the $O_{\mathcal{S}(\mathcal{A}),\alpha}$ observable dynamics.
In the presence of coherent interacting terms, $J_{xx} \neq 0$, we see a peculiar effect. For couplings $\omega_x/\kappa$ above critical the mediated field also induces persistent oscillations in the anti-symmetric sector, however these are not locked in frequency to the symmetric sector anymore. In fact, these oscillations display now beatings around $\nu_S$, as shown in Fig.(\ref{fig.btc}b). The interactions opens a gap in the synchronization of their frequencies.

\section{Perspectives}
\label{sec.conclusion}

A natural extension of our results encompasses the study of several ensembles of atoms and the resulting cooperative dynamics that may emerge closer to the many-body limit. Investigating scenarios where inhomogeneous initial conditions are introduced for $M\sim 10$ could be relevant for cavity QED experiments. In these platforms     thousands of atoms are loaded onto roughly $M\sim 10$ stacked layers (referred to as "pancakes"), where interactions occur solely through one or several cavity modes~\cite{Monika, Lev}, resulting effectively in all-to-all interactions among spins. An intriguing avenue opens by leveraging the oscillatory dynamics arising from non-uniform initial states. This approach could potentially induce self-driven instabilities within collective light-matter interfaces offering the  prospect of generating spin squeezing or non-Gaussian entanglement in systems with $U(1)$ invariance. 

Another natural extension lies in the realm of purely dissipative dynamics. While our current work has concentrated on the interplay between Hamiltonian and non-unitary processes,   structured dissipation can lead to counter-intuitive effects distinct from their unitary counterparts. For instance, a captivating direction involves investigating synthetic~\cite{seth1,seth2, Jamir_2022} or natural~\cite{Zoubi2010,chang2017} correlated emission processes initiated with inhomogeneous initial states in the semiclassical limit, which remain largely unexplored.

In summary, our results underscore that despite the well-established nature of all-to-all interacting systems, substantial opportunities remain by exploring the whole manifold of initial conditions, opening to potential new opportunity in manipulating the dynamics of quantum information, even when dynamics are collective and far from the deep quantum regime. 

\section{Acknowledgements}
We acknowledge valuable discussions with Jonathan
Keeling, Benjamin Lev, Brendan Patrick Marsh, David
Atri Schuller, Ronen Kroeze and Zhendong Zhang.
J. Marino acknowledges support by the Deutsche Forschungsgemeinschaft (DFG, German Research Foundation) – Project-ID 429529648 – TRR 306 QuCoL-
iMa (“Quantum Cooperativity of Light and Matter”), and by the QuantERA II Programme that has received funding from the European Union’s Horizon 2020 research and innovation programme under Grant Agreement No 101017733 (’QuSiED’) and by the DFG (project number 499037529). F.I. acknowledges financial support from Alexander von Humboldt foundation and the 
Brazilian funding agencies CAPES, CNPQ, and FAPERJ (Grants No. 308205/2019-7, No. E-26/211.318/2019, No. 151064/2022-9, and No. E-26/201.365/2022). D. Chang acknowledges support from the European Union, under European Research Council grant agreement No 101002107 (NEWSPIN); the Government of Spain under the Severo Ochoa Grant CEX2019-000910-S [MCIN/AEI/10.13039/501100011033]); QuantERA II project QuSiED, co-funded by the European Union Horizon 2020 research and innovation programme (No 101017733) and the Government of Spain (European Union NextGenerationEU/PRTR PCI2022-132945 funded by MCIN/AEI/10.13039/501100011033); Generalitat de Catalunya (CERCA program and AGAUR Project No. 2021 SGR 01442); Fundaci\'{o} Cellex, and Fundaci\'{o} Mir-Puig.

\end{document}